\documentstyle[11pt,paspconf]{article}

\begin{document}

\title{Compact Radio Cores in the Galactic Center and
Elsewhere$^1$}

\author{Heino Falcke}
\affil{Department of Astronomy, University of Maryland,
    College Park, MD 20742-2421, USA (hfalcke{@}astro.umd.edu)}




\begin{abstract}
Compact radio cores are not only common in radio galaxies and quasars
but also in many nearby galaxies with low-active, supermassive black
holes. One famous example is the Galactic Center source Sgr A*. Recent
studies of proper motions and radial velocities of stars in the inner
parsec of the Galaxy convincingly demonstrate the presence of a
compact dark mass of $2.5\cdot 10^6 M_\odot$ in the nucleus of the
Milky Way. Millimeter VLBI and submm observations of Sgr A* thus probe
a region of only a few Schwarzschild radii in diameter. In this paper
I will review our current theoretical and observational knowledge of
this source and compare it to some famous LINER galaxies like NGC
4258, NGC 3079, and NGC 6500. In all cases these radio cores can be well
explained by a standard AGN jet model, and, with the exception of Sgr
A*, large scale outflows are observed that have powers comparable to
those inferred from the radio cores. Recent VLBI observations of
radio-weak quasars and HST observations of Seyfert galaxies indicate
that these AGN also produce powerful jets which, however, have
relatively less luminous radio cores than radio-loud quasars and the
LINERs discussed here. Therefore, jets and compact radio cores appear
to be natural constituents of an AGN, but the reason why apparently
some jets are radio-loud and others not remains a mystery.
\end{abstract}


\keywords{Galactic Center, Compact Radio Cores, Sgr A*, LINER Galaxies,
Seyfert Galaxies, Radio-weak Quasars}


\section{Introduction}

\footnotetext[1]{Invited Review, to appear in: IAU Coll. 163,
``Accretion Phenomena and Related Outflows'', Wickramasinghe D., 
Ferrario L., Bicknell G. (eds.), PASP Conf. Proc.}
When talking about compact radio cores the first thing that comes to
ones mind are usually the flat-spectrum radio cores in quasars and
radiogalaxies to which a substantial fraction of VLBI time and many
theoretical papers have been dedicated. Those cores are generally
believed to be the bases of relativistic radio jets produced by the
central engine of an AGN. However, these cores are probably only the
tip of an iceberg, as the largest number of AGN reside in low-active
and less prominent galactic nuclei. Large surveys (Wrobel \& Heeschen
1984, Roy et al.~1994, Slee et al.~1994, Sadler et al.~1995) have
revealed that quite a substantial number of nearby galaxies also show
compact radio cores in their nuclei. As the presence of radio cores
seems to be correlated with optical emission lines (O'Connell \&
Dressel 1978) and since one third of all galaxies show evidence for
optical activity (Ho et al.~1995), we can expect to find compact radio
cores almost everywhere in the universe. One of the most prominent
examples of such a radio core is of course Sgr A* in the Galactic
Center.

While over the years quite a few observational programs have tackled
compact radio nuclei in nearby galaxies and low-power AGN, there have
been only a few detailed discussions of their physical
nature. Therefore, in the first part of this paper I will summarize
the current observational and theoretical status of Sgr A*, which will
be basically an update of my earlier reviews (Falcke 1996a\&b), and
then move on to discuss in detail examples of radio cores and jets in
prominent nearby galaxies like NGC 4258, NGC 3079, or NGC 6500. Finally,
I want to briefly discuss recent findings for Seyfert galaxies and
radio-weak quasars, which suggest that there jets play an even larger
r\^ole than initially thought.

\section{Sgr A* --- Observations}
\subsection{Mass}
The largest progress in the Galactic Center has been made in the
determination of the mass of Sgr A*. Almost one decade ago Genzel \&
Townes (1987) published their often referenced ``enclosed mass''
diagram which suggested the presence of a central dark mass of
$2\cdot10^6 M_\odot$ in the center of the Galaxy and since then it has
always been discussed how significant this result is (e.g.~Sellgren
1989, Saha et al.~1996). But now K-band spectroscopy of the GC region
(Haller et al.~1996, Genzel et al.~1996) has confirm the presence of a
dark mass of $M_\bullet=2.5\cdot10^6 M_\odot$ within 0.1 pc of Sgr A*
with a $>6\sigma$ significance. Even more fascinating is the detection
of proper motion for some of the stars in the very vicinity of Sgr A*
with Speckle imaging (Eckart \& Genzel 1996a\&b). This combined with
the velocity dispersions determines the velocity field of the stars in
three dimensions and one infers a mass density of $6.5\cdot10^9
M_\odot {\rm pc}^{-3}$. The accuracy of those Speckle images was
impressively confirmed at a recent conference by an independent K-band
speckle image with 50 mas resolution obtained with the Keck telescope
(Klein et al.~1996). The race for the most accurate mass determination
has just begun: with more 8 and 10 meter telescopes doing NIR Speckle
imaging and HST dedicating a lot of its NICMOS time to the Galactic
Center (Rieke 1996) we can expect that these results will improve even
more dramatically. Very interesting in this respect is the
announcement of a detection of a high proper motion star ($\sim1500$
km/sec) just a fraction of an arcsecond away from Sgr A* (Eckart \&
Genzel 1996a\&b). For the cluster of faint NIR stars immediately
surrounding Sgr A* the Keplerian motion around the black hole at a
distance of 0.1 arcsecond is 1600 km/sec and one would expect to see a
proper motion of the order 40 mas/yr for the nearest stars. This
should be well detectable within the very near future and in the
extreme limit could push the central dark mass density to almost
$10^{13} M_\odot/{\rm pc}^3$. The disovery of a cusp in the stellar
distribution could also increase the lower limit on the mass of the
radio source Sgr A* derived from the absence of its proper motion
(Backer 1996) from several 100 $M_\odot$ to something more like $1000
M_\odot$. Given that prospect and the already stunning success NIR
Speckle imaging had in the Galactic Center, it is really surprising
that so little time and attention is given to NIR Speckle imaging
programs with Keck --- if that would have different the matter could
have been settled by now.

\subsection{Spectrum}

While the mass determination for the Galactic Center is getting more
and more exciting every year, the spectrum of Sgr A* is getting more
and more disappointing. This source has a long history of detections
at wavelengths other than the radio, none of which survived very
long. The latest frustration stems from efforts to accurately align
the radio and the optical reference frame using Masers associated with
stars (Menten et al.~1996). This work seems to indicate that none of
the sources detected in NIR Speckle images coincide with the radio
source Sgr A*. There is a claim now of a detection of a source at
8.7$\mu$m (Stolovy et al.~1996) within 0.3\arcsec of Sgr A* which was
not seen in earlier observations (Gezari 1996). This needs to be --
and will be -- checked by further observations with higher resolutions
(e.g.~with Keck, Morris et al., in prep.). Except for a possible weak
ROSAT detection (Predehl \& Tr\"umper 1994), there is currently no
unambiguous evidence for a detection of Sgr A* at any higher frequency
(Maeda et al.~1996).

Hence, the only informations we can use right now are radio data.
There is some confusion in the current literature about the actual
spectrum of Sgr A*. Duschl \& Lesch (1994) compiled an average
spectrum from the literature and claimed a $\nu^{1/3}$ spectrum
indicative of optically thin emission from monoenergetic electrons.

However, in simultaneous multi-frequency VLA observations (Wright \&
Backer 1993) the actual spectrum was bumpy and the spectral index
varied between $\alpha=0.19-0.34$ ($S_{\nu}\propto\nu^\alpha$). Morris
\& Serabyn (1996) published a more recent spectrum that was a smooth
powerlaw with $\alpha=0.25$ below 100 GHz. Moreover, in 24
simultaneous observations at 2.7 and 8 GHz between 1976 and 1978 the
spectral index varied between $\alpha_{2.7/8}=0.08$ and
$\alpha_{2.7/8}=0.56$ with a mean value of
$\alpha_{2.7/8}=0.28\pm0.01$ (Brown \& Lo 1982). This data is
apparently affected (steepened) by the low-frequency turnover of the
spectrum around 1 GHz which seems to be variable in frequency with a
timescale of several years: in 1976 the average 2.7 to 8 GHz spectral
index was $\alpha_{2.7/8}=0.34$ wile in the consecutive two years it
was significantly lower at $\alpha_{2.7/8}=0.23$. More recent
observations with the GBI showed a steady decline at low frequencies
with a 2.7 GHz flux now around 100 mJy and a basically unchanged 8 GHz
flux (T. Minter, priv.~comm.) yielding a spectral index around 1.5. At
shorter wavelengths the spectral index is markedly flatter. The
simultaneous VLA observations of Sgr A* in 1990/91 by Zhao et
al.~(1992) indicate an average spectral index of
$\alpha_{8.4/15}\simeq0.22$ between 8.4 and 15 GHz and
$\alpha_{15/22}\simeq0.15$ between 15 and 22.5 GHz. The spectrum
changes once more if one looks at higher frequencies: the average flux
at 22.5 GHz is 1 Jy (Zhao et al.~1992) while it was 2.9$\pm0.3$ Jy at
230 GHz in the years 1987-1994 (Zylka et al.~1995) yielding
$\alpha_{22/230}=0.45\pm0.5$.

So why am I mentioning all these number? Mainly because we had heated
debates in recent years about this spectrum. As Sgr A* may be an
archetype for compact radio cores in other galaxies, extreme care has to
go into the interpretation of its spectrum --- most likely any other
radio core will have much less information available, and any error we
make here will propagate into endless space.

In conclusion, if we want to describe the radio spectrum of Sgr A* we
have to deal with at least four regimes: 1) a slowly variable low
frequency turnover somewhere below 5 GHz, 2) a fairly flat powerlaw in
the cm regime with a spectral index around 0.2, 3) an excess towards
the sub-mm with much larger spectral index, and 4) a cut-off towards
the IR (see Morris \& Serabyn 1996). Therefore it is {\it not}
possible to describe the spectrum from cm to submm wavelengths with a
single powerlaw.

\subsection{Size}
Concerning the size of Sgr A*, there have been no new developments in
the last years. The smallest size found with VLBI so far is $<1$ AU at
3 mm (Rogers et al.~1994, Doeleman et al.~1995) consistent with
scatter broadening even at these frequencies. So far the weak extended
component found by Krichbaum et al.~(1993) has not been seen again and
recent measurements of the different groups at 3 mm are in reasonable
agreement. This is not necessarily surprising as Sgr A* is strongly
variable with frequent bursts on a timescale of months (Zhao et
al.~1992) which in AGN are often associated with the expulsions of
blobs. Continued VLBI monitoring at 7 mm and 3 mm, possibly linked to
flux monitoring, is therefore necessary and further 3 and even 1 mm
VLBI experiments are already being evaluated (Krichbaum et al.~, in
prep.).

\subsection{Imaging a black hole with submm VLBI}
As pointed out in Falcke (1996a,b), the existence of a submm excess in
the Sgr A* spectrum (see above) indicates that synchrotron-self
absorption is present between 100 and 300 GHz and this requires a
source size of $\sim$ 2 Schwarzschildradii ($R_{\rm s}$). For
comparison, I note that a Schwarzschild black hole has an innermost,
marginable stable orbit of 3 $R_{\rm s}$ while an extreme Kerr black
hole has 0.5 $R_{\rm s}$. Hence, with the increasingly more accurate
mass and spectrum determination of Sgr A* we might eventually be able
to give lower limits for the spin of the black hole.  Even more
exciting is the prospect that we might actually be able to image the
horizon of the black hole and, in contrast to his pessimism, see some
of the effects R. Wehrse discussed during this conference. A black
hole embedded in front of a luminous object (or embedded in a
transparent synchrotron-emitting plasma blob) would indeed appear as a
black hole. The shape of this ``hole'' was calculated by Bardeen
(1973, Fig. 6) which in the case of Sgr A* should have a diameter of
25 micro-arcsecond. Observations with such a beamsize are in principle
possible with global mm- and submm-VLBI and hence Sgr A* should serve
as a strong motivation for developing these techniques to their full
potential.

\section{Sgr A* --- Theory}

\subsection{Low- and high-frequency cut-off}
While describing the spectrum of Sgr A* is difficult, modeling it is
even more so. The low-frequency turn-over can, however, be understood
fairly well by free-free absorption by the H {\sc II} region Sgr A
West in which Sgr A* is presumably embedded (see e.g.~Beckert et
al.~1996). Of course an intrinsic cut-off, e.g.~due to
self-absorption, can not be excluded, but the effects of free-free
absorption are readily seen already at larger scales (Pedlar et
al.~1989). The variability time scale of the cut-off frequency of a
few years, requires that a substantial amount of the absorption
happens very close to the center at a distance of roughly $10^{16}$ cm
(assuming Kepler velocities).  This scale is approximately the scale
of the closest NIR stars and also the scale derived for the location
of the refractive interstellar scintillation (RISS) which Zhao et
al.~(1989) suggested is responsible for the variability of Sgr A* at
lower frequencies. But it is unlikely that the same material also
produces the scatter-broadening of Sgr A*, as one sees similar amounts
of broadening in OH Masers throughout the central 100 pc (Frail et
al.~1994).

The high-frequency cut-off of Sgr A* in the sub-mm is usually
explained by an absence of high-energy electrons, either because of a
thermal (Melia 1992, Narayan et al.~1995) or a quasi-monoenergetic
electron distribution, the latter being due to either magnetic
reconnection (Duschl \& Lesch 1994), or as a consequence of hadronic
processes (Falcke 1996a).

\subsection{Accretion models}

The first coherent attempt to understand Sgr A* was made by Rees
(1982) where he suggested that Sgr A* is a black hole accreting
directly from the ISM. He pointed out that, as the infall timescale is
short compared to the cooling timescale, a hot ion supported torus
should form around Sgr A* and that the synchrotron radiation would be
produced by hot thermal electrons. The only crudely estimated spectral
index for this model was between $\alpha=0.4$ and $\alpha=1.3$ and
hence far too high.

In recent years quite a few modern variations of the Rees model have
appeared in the literature. Melia (1992, 1994, \& 1996) and Ozernoy
(1992) explored the possibility that Sgr A* accretes from nearby
stellar sources rather than from the ISM. However, as indicated in
Falcke (1996a) to properly determine the wind accretion rate one will
have to take the proper position (see Melia 1996) and angular velocity
of the stars into account, this is especially important, as Eckart \&
Genzel (1996) and Genzel et al.~(1996) report that the He {\sc I}
stars seem to form a counterrotating entity.

The Melia model and standard accretion models, like the ``starving
disk'' (Falcke et al.~1993a, Falcke \& Heinrich 1994), which have also
been proposed now face some serious problems with the NIR fluxes: the
observational limits are getting too low. A remedy would be an
obscuring (micro-)torus around Sgr A*, and there are some indications
in the putative ROSAT counterpart to Sgr A* that there indeed is
intrinsic absorption (Predehl \& Tr\"umper 1994, Predehl \& Zinnecker
1996) in Sgr A*. Despite this weak sign of hope, Melia and I have
joined forces to save our respective models from the advent of even
more sensitive observations and have suggested that a fossil accretion
disk in Sgr A* could capture a large fraction of the incoming wind
already at larger radii. This could reduce the amount of optical and
NIR emission substantially (Falcke \& Melia 1996).

As an alternative model Narayan et al.~(1995) suggested an advection
dominated disk model for the Galactic Center which has a concept very
similar to the Rees model with wind accretion. The main plus of this
model is that, because $>99.9\%$ of the energy is swallowed by the
black hole, it predicts very low NIR and UV luminosities (but with
further decreasing limits on Sgr A* even these could become too
high). On the other hand, the model only fits the submm bump but not
the cm-to-mm radio spectrum and the assumed black hole mass of
$M_\bullet=5\cdot10^5M_\odot$ is a factor 5 too small. The mass is a
critical parameter for the submm spectrum and it is quite likely, that
for a higher mass the peak in the spectrum would shift to lower
frequencies, making the fit even worse.

\subsection{Jet-models for Sgr A*}
Early on in the discussion, Reynolds \& McKee (1980) argued that Sgr
A* must be an outflow because its energy density $u$ is so large that it
cannot be constrained by gravity. For gravitational confinement of a
plasma bubble with radius $R$, we
would require
\begin{equation}
GM_{\bullet}\rho/R > u.
\end{equation}
Synchrotron theory and equipartition arguments give us a {\it lower}
limit for the energy density in the plasma for a fluxdensity $S_\nu$
at a frequency $\nu$. Using the above equation this translates into
an upper limit for the size of Sgr A*
\begin{equation}
R < 3.5\cdot10^{13}\,{\rm cm}\;
(S_\nu/{\rm Jy})^{1/10} (\nu/5 {\rm GHz})^{-3/10}
(M_\bullet/2.5\cdot 10^6 M_\odot)^{7/10} 
\end{equation}
and a lower limit for the brightness temperature
\begin{equation}
T_{\rm b}>2.4\cdot10^{12}\,{\rm K}\; 
(S_\nu/{\rm Jy})^{4/5}(\nu/5 {\rm GHz})^{-7/5}
(M_\bullet/2.5\cdot 10^6 M_\odot)^{-7/5}.
\end{equation}
Such a high $T_{\rm b}$ would violate the Compton limit and therefore
Sgr A* can not be gravitationally bound. Some recent models violate
the condition in Eq.~1 by an order of magnitude and thus would require
well ordered magnetic fields to confine the plasma, but that would
also imply large polarization which is not seen.

Therefore, we have developed the jet/disk symbiosis model for the
Galactic Center and other compact radio cores (Falcke et al.~1993b,
Falcke \& Biermann 1995, Falcke 1996c). The basic idea is that a jet
flow is most likely coupled to an accretion disk which determines the
jet power. For black holes the escape speed is relativistic and it is
likely that the sound speed as well as the bulk velocity of the flow
is also relativistic. Given a certain fraction of the accretion power
which is channeled into the jet (here 50\% of the dissipated energy)
and assuming an equipartition situation, one can calculate the
synchrotron radiation produced by the jet, yielding a flat to inverted
spectrum. For many sources, e.g.~like radio-loud quasars (Falcke et
al.~1995) and Sgr A*, all equipartition factors have to be close to
unity to get the very efficiently produced radio emission
observed. Geometry, velocity field, and energy densities are
self-consistently determined by assuming a free, supersonic flow which
is slowly accelerated by its own pressure gradient.

For the Galactic Center it turned out, that a very small accretion
rate would be enough to produce the 1 Jy source Sgr A*, with a size
small enough not to be resolved with the current VLBI experiments.
Alas, as pointed out above, Sgr A* is not an ideal source anymore to
test the validity of this model --- too many parameters are still
unconstrained. Therefore, we have applied this model to other galactic
nuclei, like M81, where it perfectly fits the observed data. In the
following, I want to discuss several other Galaxies with compact radio
cores, that might be directly relevant to the discussion in the
Galactic Center.

\section{Jets and compact radio cores in nearby Galaxies}
\subsection{NGC 4258}

One of the most famous galaxies in recent years was NGC 4258. VLBI
observations of H$_2$O masers in this LINER galaxy have revealed a
thin Keplerian molecular disk around the nucleus with a central dark
mass of $M_\bullet=3.6\cdot10^7M_\odot$ (Miyoshi et al.~1995). Earlier
VLA observations (Turner \& Ho 1994) showed a three mJy, compact radio
core with inverted spectrum. At this conference, Herrnstein et
al.~(1996) presented VLBA observations of NGC 4258 at 1.3 cm where
they found a variable continuum source with a flux of 1-6 mJy which
was 0.4-0.8 mas {\it offset} from the dynamical center of the masing
disk in the direction of the rotation axis. Once more this galaxy
proves to be an important laboratory because it is the only case where
we are able to determine the dynamical center of a disk with respect
to the radio core. NGC 4258 seems to confirm the jet model in which,
because of its conical geometry, the innermost part of the jet close
to the center becomes synchrotron self-absorbed. If the radio core
were for example an accretion disk atmosphere like in some Sgr~A*
models, one would expect to see radio emission centered on the
nucleus.

We can now apply the jet-disk symbiosis model as used for Sgr A* and
M81* (Falcke 1996c) without any modifications to NGC 4258. As the
inclination angle of the disk is already fixed at $i\sim83^\circ$ by
the VLBA maser observations, the model has only two adjustable
parameters, the accretion disk luminosity and the typical Lorentz
factor $\gamma_{\rm e}$ of the radiating electrons, for two observed
data points (offset and flux).  It predicts a flux of 3 mJy and an
offset of 0.4 mas if the accretion disk luminosity of NGC 4258 is
$L_{\rm disk}\sim4\cdot10^{41}$ erg/sec and $\gamma_{\rm
e}\sim700$. The latter value is relatively large (e.g.~three times
larger than estimated for M81*), but would become lower if the
observed offset is overestimated, for example because NGC 4258 expels
radio blobs more frequently than it was observed. On the other hand,
the accretion disk luminosity -- here solely derived from the radio
properties alone -- is well within current estimates: St\"uwe et
al.~(1992), trying to fit observed line ratios, postulate a hidden UV
source of $L\sim3\cdot10^{41}$ erg/sec, and Wilkes et al.~(1995)
estimate $L\sim10^{42}$ erg/sec from light scattered off dust. This
luminosity is also consistent with the accretion rated estimated by
Neufeld \& Maloney (1995) for the molecular disk. Hence, we can in
principle construct a self-consistent, quantitative model, where the
accretion rate measured in the masing disk, feeds a black hole and an
inner disk that produces x-rays, UV emission, and a radio jet plus
counterjet, which in turn illuminate the disk and provide seed
photons for the masers. Consequently, there is currently no need for
an advection dominated disk in NGC 4258 as proposed by Lasota et
al.~(1996). The jet/disk model can also explain the large scale
emission-line jet in NGC 4258, since its kinetic power is of the order
$10^{42}$ erg/sec --- as derived from the mass ($2\cdot10^6 M_\odot$)
and velocity ($\sim2000$km/sec) of the emission-line gas (Cecil et
al.~1995). One of the conclusions of the jet/disk symbiosis model
always was that jet and disk have comparable powers and NGC 4258 seems
to confirm this.

\subsection{NGC 3079}
We can further check the same model with yet another Megamaser, namely
the LINER NGC 3079. This highly inclined galaxy ($i=83^\circ$,
$D=18$Mpc) has huge radio and optical lobes (superbubbles)
perpendicular to the plane of the galaxy with a power of the order
$\dot E_{\rm kin}\sim6\cdot10^{42}$ erg/sec (Veilleux et al.~1994),
however, at present it is not clear whether the lobes are due to
starbursts or a central engine. The nuclear IR luminosity of NGC 3079
is $\sim3\cdot10^{43}$ erg/sec and we assume that the luminosity of
the AGN is of similar order. In the very nucleus we find a compact,
milliarcsecond radio core with a flux of $\sim65$ mJy. And indeed for
an $L_{\rm disk}\sim3\cdot10^{43}$erg/sec the jet/disk symbiosis model
predicts the right size and flux of the compact radio core, if the
electron Lorentz factor is $\gamma_{\rm e}=275$ and the inclination
$\sim75^\circ$. The high inclination agrees well with the high
inclination of the galaxy (which, however, need not always be related)
and the fact that we see maser emission. Most importantly, this means
that the relatively high luminosity of the radio core requires a
powerful central engine and a powerful radio jet which in fact would
be sufficient to drive the lobes without the help of a
starburst-driven superwind. In summary, this galaxy could in many
details be very similar to NGC 4258 with one important difference: if
we compare the ratio $R_{\rm l/c}$ between the lobe and the core radio
flux in both galaxies, we find that NGC 4258 has $R_{\rm l/c}\sim200$,
while NGC 3079 has only $R_{\rm l/c}\sim5$. The best explanation for
this discrepancy is the different orientation of the lobes: in NGC 4258
the jet has to go through the disk of the galaxy and the ISM, which
may increase the radiative efficiency, while in NGC 3079 the lobes go
out of the plane of the galaxy with presumably much less interaction.

\subsection{NGC 6500}
One important assumption in the modeling so far was that all these
radio jets start out with relativistic speeds and bulk Lorentz factors
of $\gamma_{\rm j}\sim2-3$ because mildly relativistic speeds are
required if the jets want to escape from the vicinity of the black
hole. A necessary consequence would then be relativistic boosting in
systems seen face on. One such case could indeed be NGC 6500
($i=34^\circ$, $D=63$ Mpc), which has a 200 mJy radio core with a size
of $\sim3$ mas (Jones et al.~1981) and large-scale lobes as in NGC
3079 (Unger et al.~1989). The nuclear luminosity of this galaxy cannot
be much larger than in NGC 3079, since its IR luminosity is comparable
to NGC 4258, yet its radio core is much brighter. Nor surprisingly,
the jet/disk model predicts that the inclination angle for the radio
jet has to be around $25-30^\circ$ in order to explain the bright
radio core. This would mean, that we are looking right into the
boosting cone of the jet and one could consider this galaxy a
low-luminosity equivalent to a Blazar.

\section{Seyferts and radio-weak quasars}
So far we have discussed only LINER galaxies with compact radio
cores. But what is the situation in Seyfert galaxies? Surprisingly,
only 10\% of Seyfert galaxies show flat spectrum cores (deBruyn \&
Wilson 1978), and Sadler et al.~(1995) found that the pc scale cores
seen in many Seyfert galaxies probably have a steep spectrum. Yet,
many Seyfert galaxies apparently have large-scale jets (Ulvestad \&
Wilson 1989, and refs.~therein), and one of the important results of
HST for Seyfert galaxies is that the emission-line region (NLR) of
Seyferts can be substantially modified by the impact of the radio
jet. The best examples are perhaps MK 573 (e.g.~Capetti et al.~1996)
and ESO 428-G14 (Falcke et al.~1996b). The former has a NLR with
bow-shock like features around radio hot-spots, while the NLR of the
latter consists of thin narrow strands of emission line gas, which
apparently wrap around a radio jet and form a figure eight on one
side, suggesting helical motion. In those two cases it is obvious that
the jet dominates and shapes the emission-line gas of these Seyfert
galaxies.

Conventional wisdom suggests that Quasars are the more luminous
counterparts to Seyfert galaxies and we can ask how important are jets
therein? The answer is simple for radio-loud quasars which harbor
luminous compact radio cores and powerful relativistic jets. However,
only 10\% of all quasars are radio-loud, while the rest is radio-quiet
--- not completely quiet though, because many of them have compact, or
diffuse extended radio emission at the milli-Jansky level and
below. In a few cases Kellerman et al.~(1994) found bipolar structures
which they interpreted as radio jets. Moreover, we have recently
investigated the properties of a mysterious class of
radio-intermediate quasars (RIQ, Falcke et al.~1996a). These are
flat-spectrum core-dominated quasars, with radio luminosities
(normalized to the optical luminosity) {\it below} those of typical,
lobe-dominated steep spectrum quasars, and they do not have any
extended emission. Hence, these sources can not be relativistically
boosted radio-loud quasars (as usually assumed for core-dominated
quasars), instead, many arguments suggest that these quasars are
preferentially oriented, relativistically boosted, radio-quiet
quasars. The relatively large fraction of RIQs then also implies that
basically all radio-weak quasars have such relativistic jets in their
nuclei (but are pointing away from us).

\subsection{Are LINERs radio-loud?} In summary, there is a lot of
evidence now, that jets are integral and dynamically important parts
of Seyfert galaxies and radio-quiet quasars as well. For some reasons,
however, their radiative efficiency is much lower than in radio-loud
quasars. In this respect it is interesting to note that many of the
galaxies with prominent, compact radio cores are LINERs, like all the
galaxies discussed in this paper, and the model used here to explain
their radio-cores is exactly the same as for radio-loud
quasars. Unfortunately, there is currently no good statistical data
base to back this claim, but it might be an interesting question to
ask whether the central engine in some LINERs are the counterparts to
radio-loud quasars, while the AGN in Seyfert galaxies are the
counterparts to radio-weak quasars at lower accretion rates.

\section{Conclusions}
The evidence that the Galactic Center harbors a supermassive black
hole of mass $\sim2.5\cdot10^6$ is now close to overwhelming. The
compact radio core Sgr A* which is probably associated with this
supermassive black hole, does not appear to be a very unusual object
compared to the radio cores in the nuclei of other galaxies, unusual
are only its optical/NIR properties (or better the lack thereof).
Advection dominated disk models may explain the low radiative
efficiency, but fail to properly explain the radio spectrum, the
jet/disk model on the other hand can explain the radio spectrum fairly
well, yet predicts too much optical/NIR flux. These problems do not
necessarily exclude any of these models yet, but could mean that we
are still missing a critical (observational) piece in the Galactic
Center puzzle.

Personally, I believe that the jet model still has the most merits,
since it works well for compact radio cores in all other galaxies. The
properties of all the galaxies discussed here (Milky Way, M81, NGC
4258, NGC 3079, \& NGC 6500), i.e. the fluxes and sizes of their radio
cores, can be explained by one model with basically three input
parameters: the accretion rate, the inclination to the line of sight,
and (as the only fudge-factor), the typical Lorentz factor of the
relativistic electrons. In some of these galaxies, e.g.~like NGC 4258,
and of course all radio-loud quasars and radio galaxies, the
association between jet and compact radio core is almost
indisputable. Therefore, until demonstrated otherwise, there is no
reason to believe that in sources like Sgr A* where no large scale jet
is yet detected, the radio cores are anything but the bases of radio
jets powered by accretion onto a massive, compact object.

As Livio (1996) pointed out there is hardly any class of sources left
where accretion is not accompanied by a bipolar outflow. And as
discussed in this paper we find compact radio-cores or powerful
large-scale jets in many extragalactic sources, like LINERs, Seyferts,
and even radio-quiet quasars. This makes the jet phenomenon a truly
universal phenomenon.

\acknowledgments
I acknowledge support from NASA under grants NAGW-3268 and NAG8-1027.

%
%

\end{document}